\documentclass[aps,prc,twocolumn,showpacs]{revtex4}
\usepackage{graphicx,amsfonts,amssymb,amsbsy}
\usepackage{amsmath}
\usepackage{color,pstricks}

\usepackage{natbib}
\usepackage[colorlinks]{hyperref}
\hypersetup{linkcolor=blue,citecolor=blue}

\begin{document}

\title{Surface tension of hot and dense quark matter under strong magnetic fields}

\author{G. Lugones$^1$}
\author{A. G. Grunfeld$^{2,3}$ }
\affiliation{$^1$ Centro de Ci\^{e}ncias Naturais e Humanas, Universidade Federal do ABC, \\ Av. dos Estados 5001, CEP 09210-580, Santo Andr\'{e}, SP, Brazil}
\affiliation{ $^2$ CONICET, Rivadavia 1917, (1033) Buenos Aires, Argentina.} 
\affiliation{$^3$ Departamento de F\'\i sica, Comisi\'on Nacional de Energ\'{\i}a At\'omica, Av. Libertador 8250, (1429) Buenos Aires, Argentina}

\begin{abstract}
We study the surface tension of hot, highly magnetized three flavor quark matter droplets, focusing specifically on the thermodynamic conditions prevailing in neutron stars,  hot lepton rich protoneutron stars and neutron star mergers. 
We explore the role of temperature, baryon number density, trapped neutrinos, droplet size and magnetic fields within the multiple reflection expansion formalism (MRE), assuming that astrophysical quark matter can be described as a mixture of free Fermi gases composed by quarks $u$, $d$, $s$, electrons and neutrinos, in chemical equilibrium under weak interactions.
We find that the total surface tension is rather unaffected by the size of the drop, but is quite sensitive to the effect of baryon number density, temperature, trapped neutrinos and magnetic fields (specially above $eB \sim 5 \times 10^{-3} \mathrm{GeV}^2$).
Surface tensions parallel and transverse to the magnetic field span values up to $\sim$ 25 MeV/fm$^2$. 
For $T \lesssim 100$ MeV the surface tension is a decreasing function of temperature but above 100 MeV it  increases monotonically with $T$.
Finally, we discuss some astrophysical consequences of our results. 
\end{abstract}


\maketitle

\section{Introduction}
%
The study of the surface tension of quark matter has attracted much attention recently because of its impact on heavy ion collisions  and in astrophysics. 
In the astrophysical context, quark matter surface tension plays a key role in the understanding of neutron star (NS) interiors (see \cite{LugonesGrunfeld2017,Lugones2013} and references therein). In fact, it has been speculated that hybrid stars may contain a mixed hadron-quark phase in their interiors. In such a phase, it is assumed that the electric charge is zero globally but not locally, and therefore charged hadronic and quark matter may share a common lepton background, leading to a quark-hadron mixture extending over a wide density region of the star \cite{Glendenning2000}. The mixed phase entails a smooth variation of the energy density, leading in turn to a continuous density profile along the star. Whether the quark-hadron interface in a compact star is actually a sharp discontinuity or a wide mixed region depends crucially on the amount of surface energy needed for the formation of  drops of one phase embedded in the other  \cite{Voskresensky2003,Maruyama2007}. If the energy cost of  surface effects exceeds the gain in bulk energy, the scenario involving a sharp interface turns out to be favorable. 
The quark matter surface tension is also crucial for understanding the most external layers of strange quark stars \cite{Xia:2016guv}, where matter may fragment into a charge-separated mixture, involving positively-charged strangelets immersed in a negatively charged sea of electrons, presumably forming a crystalline solid crust \cite{Jaikumar2006}. This scenario would be favored  below some critical surface tension which is typically of the order of a few MeV/fm$^2$ \cite{Alford2006}.
However, in spite of its relevance for NS physics, the surface tension is still poorly known for quark matter.  Early calculations as well as recent calculations within different models give rather low values \cite{Berger1987,Pinto2012,doCarmo2013,Fraga2018,LugonesGrunfeld2017}, typically below $\sim  30 \, \mathrm{MeV/fm^2}$, but significantly larger results (above $\sim 100 \, \mathrm{MeV/fm^2}$) were obtained by other authors   \cite{Bugaev,Lugones2011,Lugones2013,EPJC_nuestro}.

In this work we study the surface tension of hot, highly magnetized three flavor quark matter droplets within the multiple reflection expansion (MRE) framework (for details on the MRE formalism see \cite{Balian1970,Madsen1994,Kiriyama1,Kiriyama2} and references therein).  We describe quark matter as a strongly magnetized free Fermi gas of $u$, $d$ and $s$ quarks, electrons and neutrinos, all in chemical equilibrium under weak interactions  \cite{Strick,LugonesGrunfeld2017}. We focus on the thermodynamic conditions prevailing in NSs, hot lepton rich PNSs and binary NS mergers.
NSs are believed to be born after the gravitational collapse and supernova explosion of a massive star. Initially, PNSs are very hot and lepton-rich objects, where neutrinos are temporarily trapped. Initially, the temperature may reach 40 MeV and the chemical potential of trapped neutrinos can be as high as 150 MeV. 
During the first minute of evolution the PNS evolves to form a cold NS. As neutrinos are radiated, the lepton-per-baryon content of matter goes down and the neutrino chemical potential tends to essentially zero. 
Finally, binary NS merging has recently gained attention after the LIGO/Virgo collaboration detected the  gravitational waves from the merging event GW170817 \cite{GW170817}. Numerical simulations of these events suggest that the just merged compact object may attain high temperatures \cite{Bauswein2013,Kastaun2015} (as high as $\sim 70$ MeV \cite{Radice2017}) with an expected huge amount of trapped neutrinos. As a limiting scenario we will consider here an extreme  configuration with $T= 100$ MeV and $\mu_{\nu_e} = 200$ MeV.

The article is organized as follows: in Section II we present the MRE formalism for finite size droplets immersed in a strong magnetic field. 
In Section III we include finite size effects in the thermodynamic quantities. In particular, we derive the longitudinal and transverse surface tension as well as the particle number density. We also introduce some thermodynamic constraints and present different astrophysical scenarios to be analyzed. 
In Section IV we report and discuss our results. Finally, in Section V we present our conclusions.

\section{Finite size effects in the presence of a magnetic field}

In the MRE formalism, the  modified  density of states of a finite  droplet with an arbitrary shape is given by  \cite{Balian1970,Madsen1994}
\begin{equation}
\rho_{\mathrm{MRE}}(k, m, S, V, \cdots) = 1 + \frac{2 \pi^2}{k} \frac{S}{V} f_{S}  + \cdots
\label{rho_MRE}
 \end{equation}
where $m$ is the mass of a quark or a lepton, $k$ its momentum, $S$ is the droplet's surface, $V$ its volume,  and 
\begin{equation}
f_{S}(k) = - \frac{1}{8 \pi} \left(1 - \frac{2}{\pi} \arctan \frac{k}{m} \right) 
\label{eq:fs}
\end{equation}
is  the surface contribution to the new density of states.  
As pointed out in Refs. \cite{Kiriyama1,Kiriyama2} the MRE density of states given in Eq. \eqref{rho_MRE} may become negative for small momenta. To exclude this non physical behavior,  an infrared cutoff $\Lambda$ is introduced in momentum space (see \cite{Kiriyama2} for details) where $\Lambda$ is the largest solution of the equation $\rho_{\mathrm{MRE}}(k, m, S, V, \cdots)= 0$ with respect to the momentum $k$. 

To implement the MRE formalism  we have to perform the following replacement in a generic thermodynamic integral $I$:
\begin{equation}
I \equiv \frac{1}{(2 \pi)^3} \int \cdots d^3 k \rightarrow    \frac{1}{(2 \pi)^3}  \int_{\Lambda }^{\infty} \cdots \rho_{\mathrm{MRE}}(k) \,   4 \pi k^2 dk.
\label{MRE}
\end{equation}

For matter immersed in a magnetic field $\textbf{B}$ pointing in the $z$ direction, the transverse motion of particles with electric charge $qe$ is quantized into Landau levels (LL).
The  momentum $k$ in Eqs. \eqref{rho_MRE} and \eqref{eq:fs} is given by  $k =  \sqrt{k_z^2 +  2 \nu |q e B| }$   where $\nu \geq 0$ is an integer, and the momentum integrals  in the transverse plane must be replaced by sums over the discretized levels. Thus, in the thermodynamic integrals we must use \cite{LugonesGrunfeld2017}:
\begin{equation}
I \rightarrow     \frac {|q e B|}{2 \pi^2} \sum_{\nu=0}^{\infty} \alpha_{\nu}  
\int_{\Lambda_{\nu}}^{\infty} \cdots \rho_{\mathrm{MRE}} dk_z ,
\label{MRE_with_B}
\end{equation}
where  $\Lambda_{\nu}$ are the cutoffs in the momentum along the direction of the magnetic field, which now depend on the Landau levels.  $\alpha_{\nu} =2$ for all cases except for ${\nu} = 0$, where $\alpha_{\nu} =1$.

The infrared cutoff  $\Lambda_{\nu}$ is the largest solution  of  the equation  $\rho_{\mathrm{MRE}}(k_z, m,  S, V) = 0$ with respect to the momentum $k_z$. 
Taking only the first two terms of Eq. (\ref{rho_MRE}) we obtain we obtain \cite{LugonesGrunfeld2017}
\begin{equation}
\Lambda_{\nu}  =  \sqrt{ \frac{S^2}{4V^2} x^2_{0}  -   2 \nu |q e B| } ,
\end{equation}
where $x_{0}$ is the solution of  $\lambda x = \cot x$  with $\lambda = S / (2Vm)$.  In Table \ref{cutoff} we show the values of $x_{0}$ for different quark masses and different values of $V/S$.

The thermodynamic integrals can be written in terms of the particle's energy $E = (k_z^2  + 2 \nu |q e B| + m^2)^{1/2}$ as:
\begin{eqnarray}
I \rightarrow   \frac {|q e B|}{2 \pi^2} \sum_{\nu=0}^{\infty } \alpha_{\nu}  \int_{\ell}^{\infty} \cdots \rho_{\mathrm{MRE}}(E)    \frac{E dE}{k_z(E)} ,
\end{eqnarray}
where $k(E)=(E^2-m^2)^{1/2}$,  $k_z(E)=(E^2-2\nu |qeB| - m^2)^{1/2}$, and the energy cutoff $\ell = \sqrt{ \Lambda_{\nu}^2  +   2 \nu |qeB| + m^2}$ reads:
\begin{equation}
\ell = \sqrt{ \frac{S^2}{4V^2} x^2_{0} + m^2 }.
\end{equation}
Notice that $\ell$ doesn't depend on the index $\nu$.

\begin{table}[tb]
\centering
\begin{tabular}{c c c | c}
\hline \hline
particles & $m$ [MeV]          &          $V/S$ [fm]              &   $x_{0i}$               \\
\hline  
electrons &0.511     &   10                 &  0.225631  \\   
  &0.511     &    50                &   0.487927  \\   
  &0.511     &    100                &   0.663088  \\   
  &0.511     &    $\infty$                &   $ \pi / 2$ \\   
 \hline 
quarks u, d &5     &   10                 &  0.657008  \\ 
&5     &   50                 &  1.14604   \\ 
&5      &   100                &  1.31661  \\ 
&5      &   $\infty$                 &   $\pi/2$     \\ 
\hline
quarks s &150     &   10                 &  1.47413  \\ 
&150   &   50                 &  1.5504    \\ 
&150  &   100                &  1.56053 \\ 
&150   &   $\infty$                 &  $\pi/2$      \\ 
\hline \hline 
\end{tabular}
\caption{Values of the parameter $x_0$ for different particle masses and different values of $V/S$.} \label{cutoff}
\end{table}

\section{Thermodynamic quantities at finite temperature}

\subsection{Number density}
%
The number densities can be obtained starting from \cite{Strick}
\begin{eqnarray}
n_i(T) = \frac{|q_i e B|}{2\pi^2}  \sum_{\nu=0}^{\infty} \alpha_{\nu}   \int_{0}^{\infty}  (F_{i} - {\bar F}_{i}) dk_z ,
\end{eqnarray}
where the Fermi-Dirac distribution functions for particles and antiparticles are respectively: 
\begin{eqnarray}
F_i &=& \frac{1}{e^{(E_i - \mu_i)/T} + 1} , \\
{\bar F}_i &=& \frac{1}{e^{(E_i + \mu_i)/T} + 1} ,
\end{eqnarray}
being $\mu_i$ the chemical potential for the particle species $i$.

Including the MRE density of states, we obtain:
\begin{eqnarray}
n_i  &= & \frac{|q_i e B|}{2\pi^2}  \sum_{\nu=0}^{\infty} \alpha_{\nu}   \int_{\Lambda_{i,\nu}}^{\infty} 
(F_i - {\bar F}_i) \left[1 + \frac{2 \pi^2 S}{k V} f_{S,i}\right]   dk_z  \nonumber \\
 &= & \frac{|q_i e B|}{2\pi^2}  \sum_{\nu=0}^{\infty} \alpha_{\nu}   \int_{\ell_i}^{\infty} 
(F_i - {\bar F}_i) \nonumber \\
    & & \times  \left[ 1 + \frac{2 \pi^2 S}{V} \frac{f_{S,i}(E)}{k(E)} \right]  \frac{E dE}{k_z(E)} .
\label{eq:number_density}    
\end{eqnarray}
where the energy cutoff is $\ell_i = \sqrt{S^2 x^2_{0i}/(4V^2)  + m_i^2 }$.

\subsection{Parallel pressure and surface tension}
%
A similar procedure allows writing the parallel thermodynamic potential of a magnetized quark matter drop within the MRE formalism as (see \cite{Strick,LugonesGrunfeld2017}):  
\begin{eqnarray}
- {\Omega_i^{\parallel} }
 & = & \frac { |q_i e B|}{2 \pi^2} 
\sum_{\nu=0}^{\infty} \alpha_{\nu} \int_{\Lambda_{i,\nu}}^{\infty}  \frac { (F_i + {\bar F}_i)  k_z^2 }{\sqrt{k^2 + m_i^2}}  \nonumber \\
  & &  \times  \left(V + \frac{2 \pi^2 f_{S,i}(k)}{k} \times S \right)  dk_z.
\end{eqnarray}

The latter expression can be written in the form
\begin{eqnarray}
\Omega_i^{\parallel}  = -  \Pi_i^{\parallel}  V +  \alpha_i^{\parallel}   S 
\end{eqnarray}
where $\Pi_i^{\parallel}$ is the parallel pressure within the MRE formalism and $ \alpha_i^{\parallel}$ is the parallel surface tension, i.e.,
\begin{eqnarray}
\Pi_i^{\parallel}(T) &=&  \frac{|q_i e B|}{2 \pi^2} \sum_{\nu=0}^{\infty} \alpha_{\nu}  \int_{\Lambda_{i,\nu}}^{\infty} \frac{(F_i + {\bar F}_i) k_z^2 \, dk_z}{\sqrt{k^2 + m_i^2}} \nonumber \\  
&=&  \frac{|q_i e B|}{2 \pi^2} \sum_{\nu=0}^{\infty} \alpha_{\nu}  \int_{\ell_i}^{\infty} (F_i + {\bar F}_i) k_z(E) dE \\
\alpha_i^{\parallel}(T)  &=&  - |q_i e B| \sum_{\nu=0}^{\infty} \alpha_{\nu} 
\int_{\Lambda_{i,\nu}}^{\infty}   \frac { (F_i + {\bar F}_i)  f_{S,i}  k_z^2  dk_z}{k \sqrt{k^2 + m_i^2}}  \nonumber \\
 &=&  - |q_i e  B| \sum_{\nu=0}^{\infty} \alpha_{\nu} 
\int_{\ell_i}^{\infty} (F_i + {\bar F}_i)   \frac {f_{S,i} k_z(E) dE}{k(E)} \qquad
\label{eq:parallel_surface_tension}
\end{eqnarray}
To obtain the total parallel surface tension, we have to add the contribution of all particle species, i.e. quarks $u$,  $d$, $s$ and electrons:
\begin{eqnarray}
\alpha_{\mathrm{tot}}^{\parallel} & = & \sum_{i=u, d, s, e} \alpha_{i}^{\parallel} .
\end{eqnarray}
Within the MRE formalism, neutrinos have vanishing surface tension because they are assumed to be massless; then from Eq. \eqref{eq:fs}, $f_{S, \nu_e}=0$.

\subsection{Transverse pressure and surface tension}

The  transverse thermodynamic potential of a magnetized quark matter drop reads \cite{Strick,LugonesGrunfeld2017}: 
\begin{equation}
- \Omega_i^{\perp}  =   \frac{V  |q_i e B|^2}{2 \pi^2}
\sum_{\nu=0}^{\infty} \alpha_{\nu} \nu \, \int_{\Lambda_{i,\nu}}^{\infty}  \frac{ (F_i + {\bar F}_i) \rho_{\mathrm{MRE}, i} dk_z}{\sqrt{k^2 + m_i^2}}  
\end{equation}
Again, the latter expression has the form 
\begin{equation}
\Omega_i^{\perp}  = -  \Pi_i^{\perp}  V +  \alpha_i^{\perp}   S .
\end{equation}
Therefore, the transverse  pressure $\Pi_i^{\perp}$ and the transverse surface tension $\alpha_i^{\perp}$ are given by:
\begin{eqnarray}
\Pi_i^{\perp}(T)  &=&  \frac{|q_i e B|^2}{2 \pi^2} \sum_{\nu=0}^{\infty} \alpha_{\nu} \nu \, \int_{\Lambda_{i,\nu}}^{\infty}  
\frac{ (F_i + {\bar F}_i) dk_z}{\sqrt{k^2 + m_i^2}} \nonumber \\  
&=&  \frac{|q_i e B|^2}{2 \pi^2} \sum_{\nu=0}^{\infty} \alpha_{\nu} \nu \, \int_{\ell_i}^{\infty}  
\frac{ (F_i + {\bar F}_i)}{k_z(E)} dE \\
\alpha_i^{\perp}(T) &=&  - |q_i e B|^2 \sum_{\nu=0}^{\infty} \alpha_{\nu} \nu  \int_{\Lambda_{i,\nu}}^{\infty}  \frac { (F_i + {\bar F}_i) f_{S,i} \, dk_z}{k  \sqrt{k^2 + m_i^2}} \nonumber \\
&=&  - |q_i e B|^2 \sum_{\nu=0}^{\infty} \alpha_{\nu} \nu  \int_{\ell_i}^{\infty}   \frac {(F_i + {\bar F}_i) f_{S,i}  dE}{k(E) k_z(E)}  \qquad
\label{eq:transverse_surface_tension}
\end{eqnarray}
where $i$ runs over all particle species, e.g. quarks $u, d, s$ and electrons. The total transverse surface tension is:
\begin{eqnarray}
\alpha_{\mathrm{tot}}^{\perp} & = & \sum_{i=u, d, s, e} \alpha_{i}^{\perp}  .
\end{eqnarray}
Here we also have $ \alpha_{\nu_e}^{\perp} =0$.

\subsection{Thermodynamic constrains and astrophysical scenarios}

In this work we are interested in potential astrophysical applications, therefore, we shall consider that quark matter is electrically neutral and in equilibrium under weak interactions. Chemical equilibrium is maintained by weak interactions among quarks, e.g.
$d \leftrightarrow u + e^- + \bar{\nu}_e$, $s \leftrightarrow u +
e^- + \bar{\nu}_e$, $u + d \leftrightarrow u + s$, from which we obtain the following relations between the chemical potentials:
\begin{eqnarray}
\mu_d &=& \mu_u + \mu_e - \mu_{\nu_e}, \\
\mu_s &=& \mu_d  .
\end{eqnarray}
Local charge neutrality means that:
\begin{eqnarray}
\frac{2}{3} n_u - \frac{1}{3} n_d - \frac{1}{3} n_s - n_e = 0, 
\label{charge_neutrality}
\end{eqnarray}
where the number densities are given by Eq. \eqref{eq:number_density}.

We focus here in thermodynamic conditions that are relevant for three different astrophysical scenarios: 

\begin{itemize}

\item \textit{Cold deleptonized NSs}. This is the case of most NSs a few minutes after their birth. The thermodynamic state can be characterized by a very low temperature (typically below 1 MeV) and no trapped neutrinos because their mean free path is much larger that the stellar radius. As a representative case, we shall consider here $T=1$ MeV and $\mu_{\nu_e}=0$.  This is essentially the same case already studied in our previous work \cite{LugonesGrunfeld2017} but we include it here for comparison. 

\item  \textit{Hot lepton rich PNSs}.  This is the case of NSs during the first few  minutes after their birth in core collapse supernovae. The thermodynamic state can be characterized by high temperatures (typically up to $\sim 40$ MeV) and a large amount of trapped neutrinos in the system (neutrino chemical potential $\mu_{\nu_e}$ up to $\sim 150$ MeV) \cite{Pons1999}. As a representative case we consider here $T= 30$ MeV and $\mu_{\nu_e} = 100$ MeV. 

\item \textit{Binary NS mergers} (labelled as MERGER in the figures). Very recently, the LIGO/Virgo collaboration detected the first signal of gravitational waves coming from the binary NS merger GW170817 \cite{GW170817}. Numerical simulations of these events suggest that the just merged compact object may attain very high temperatures (up to 70 MeV \cite{Radice2017}) and a huge amount of trapped neutrinos can be expected. As a limiting case we consider here $T= 100$ MeV and $\mu_{\nu_e} = 200$ MeV. 

\end{itemize}

\begin{figure}[tb]
\includegraphics[angle=0,scale=0.37]{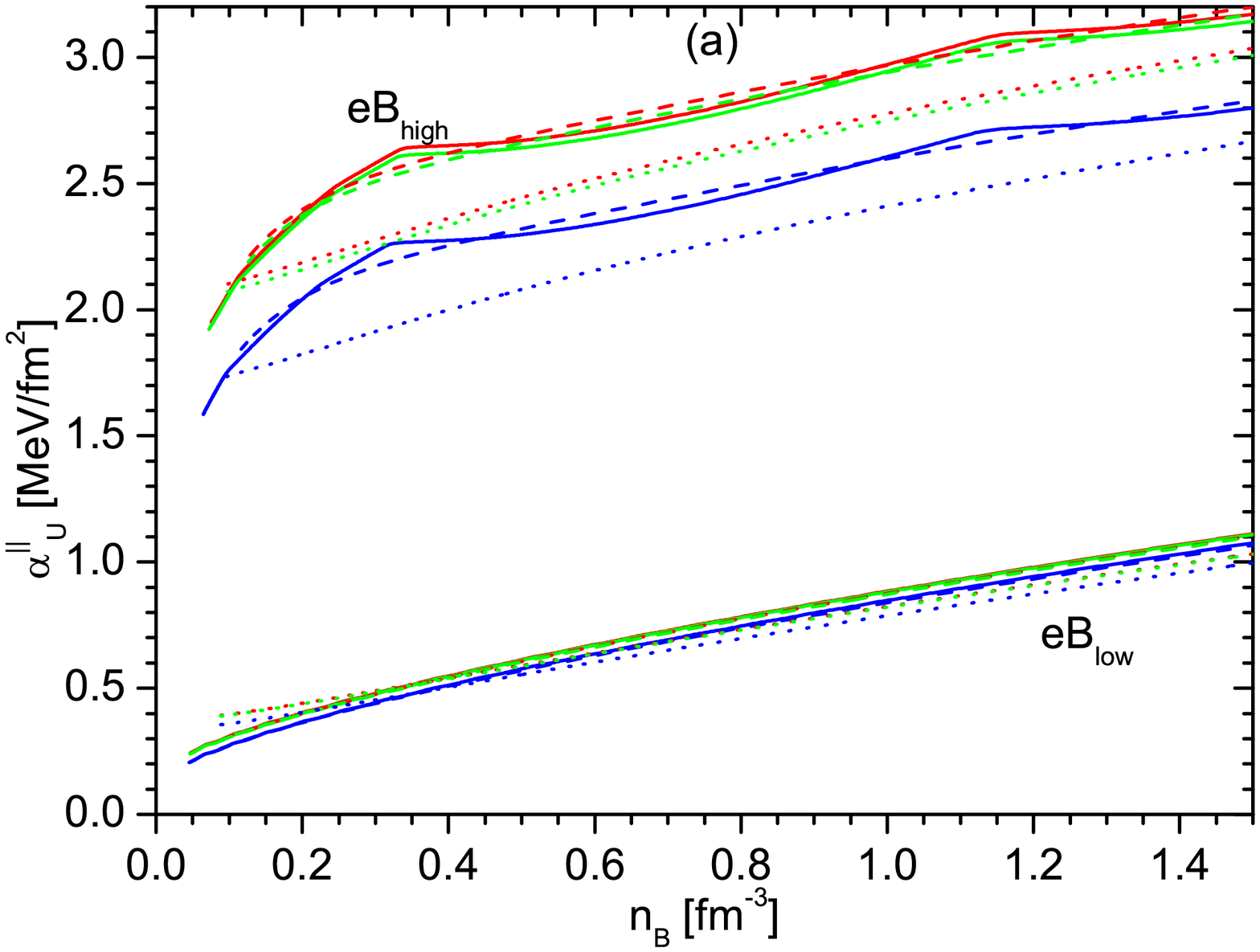}
\includegraphics[angle=0,scale=0.37]{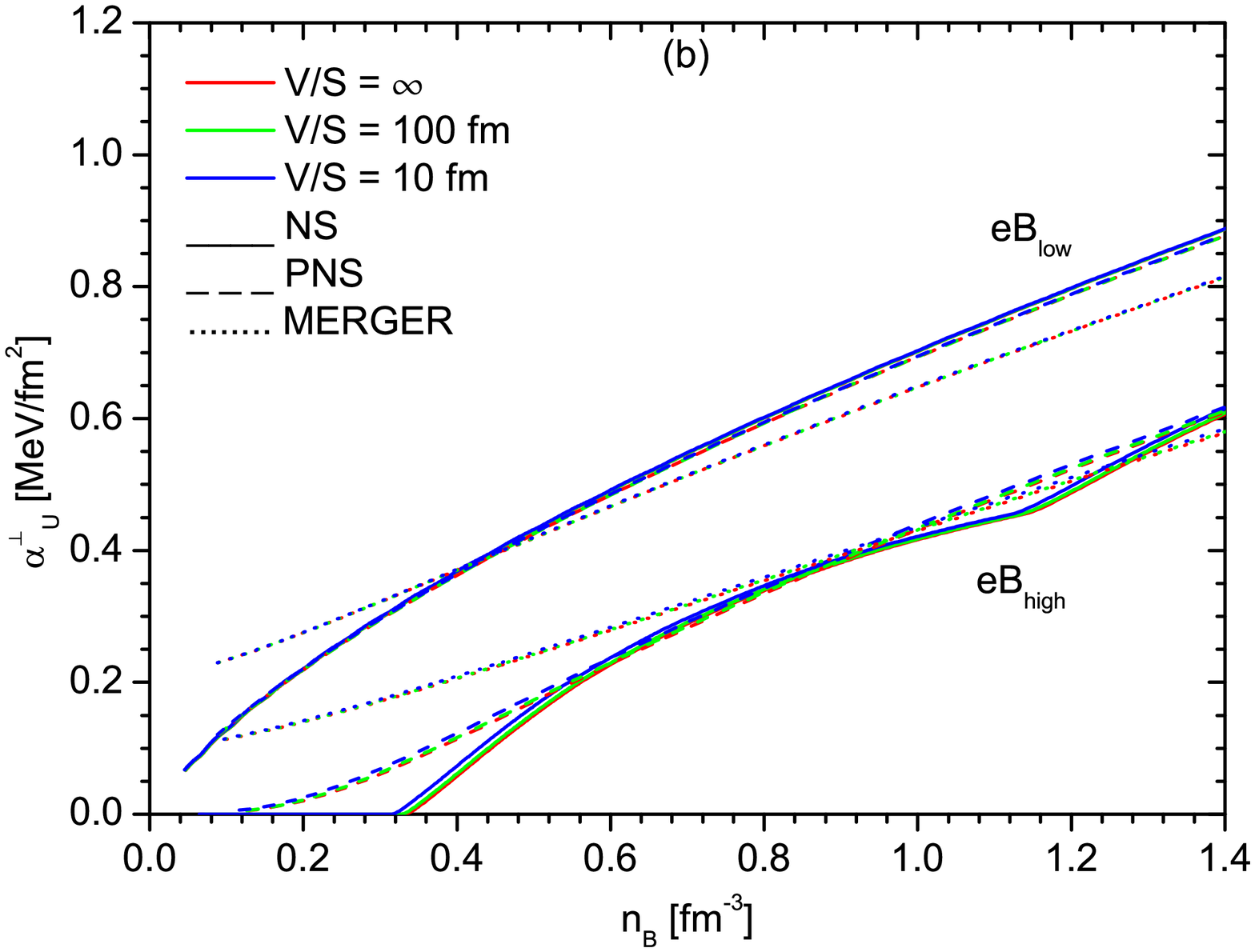}
\caption{Surface tension for quarks $u$ in the parallel direction (panel (a)) and in the transverse direction  (panel (b)). The results are shown for drops with $V/S$ = 10 fm, 100 fm and  for the bulk limit with $V/S = \infty$. The magnetic field intensities are $eB_{low} = 5 \times 10^{-3} $ GeV$^2$ and $eB_{high} = 5 \times 10^{-2} $ GeV$^2$. We considered the three different astrophysical scenarios described in the text.} 
\label{fig:1}
\end{figure}

\begin{figure}[tb]
\includegraphics[angle=0,scale=0.37]{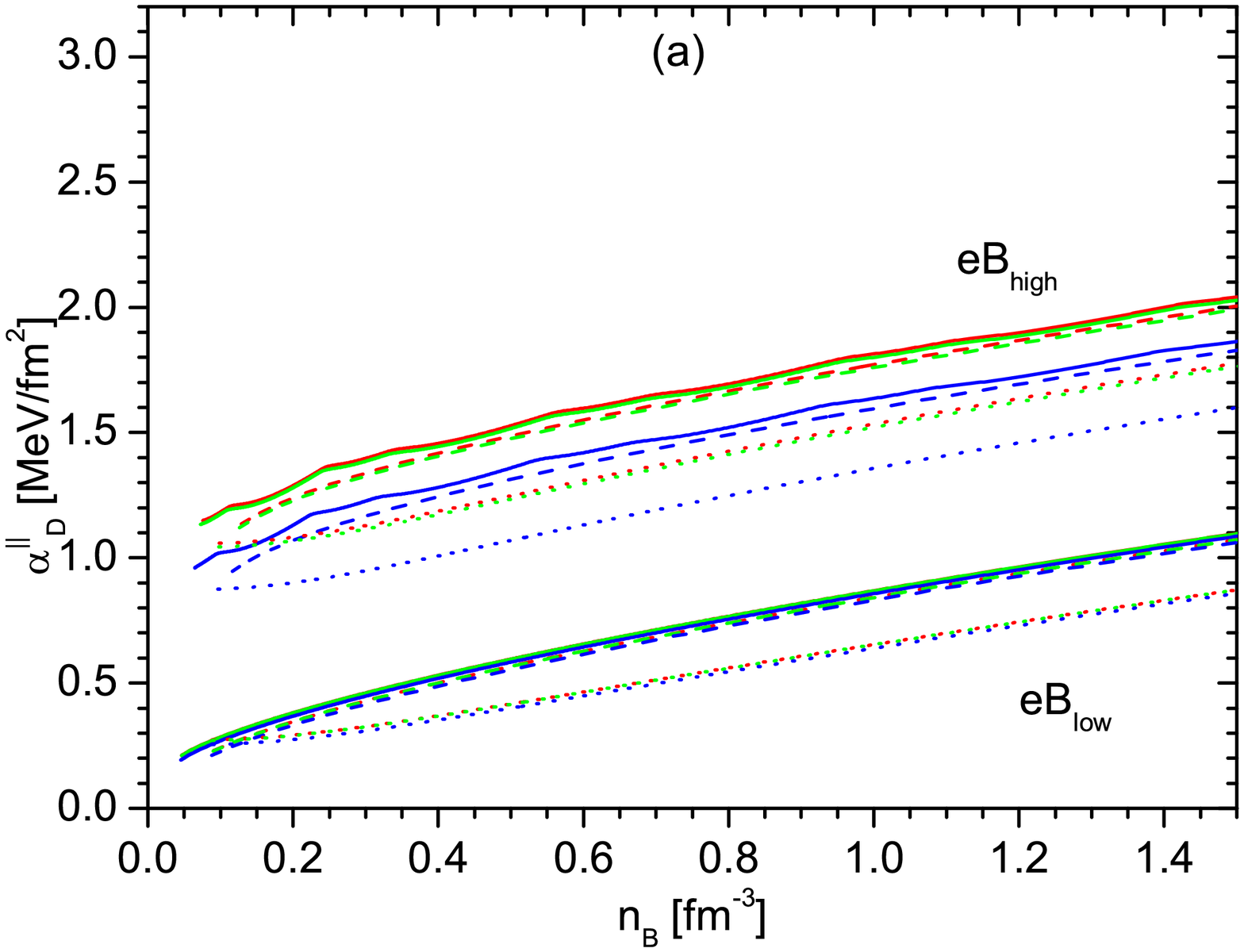}
\includegraphics[angle=0,scale=0.37]{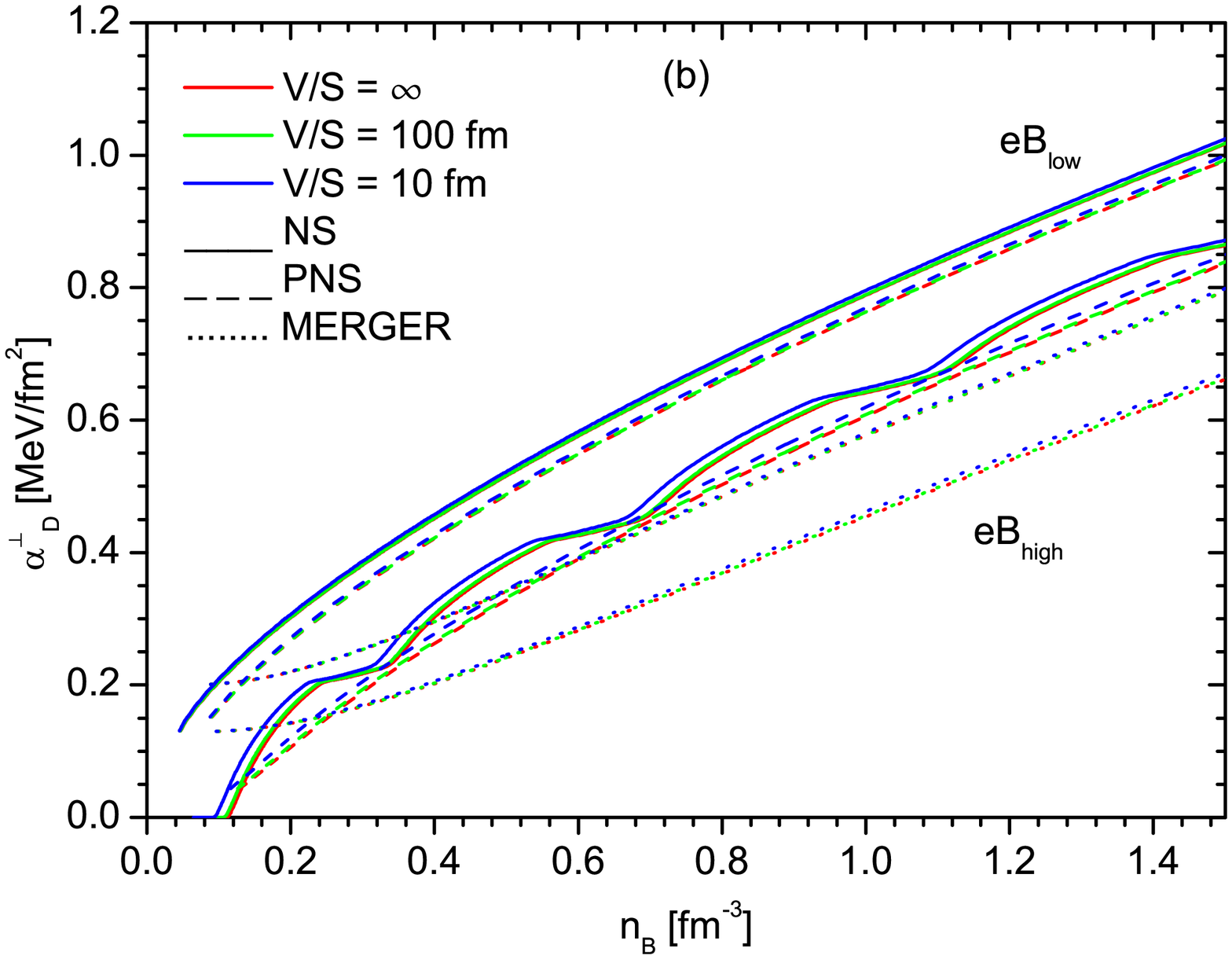}
\caption{Same as in Fig. \ref{fig:1} but for quarks $d$.} 
\label{fig:2}
\end{figure}

\begin{figure}[tb]
\includegraphics[angle=0,scale=0.37]{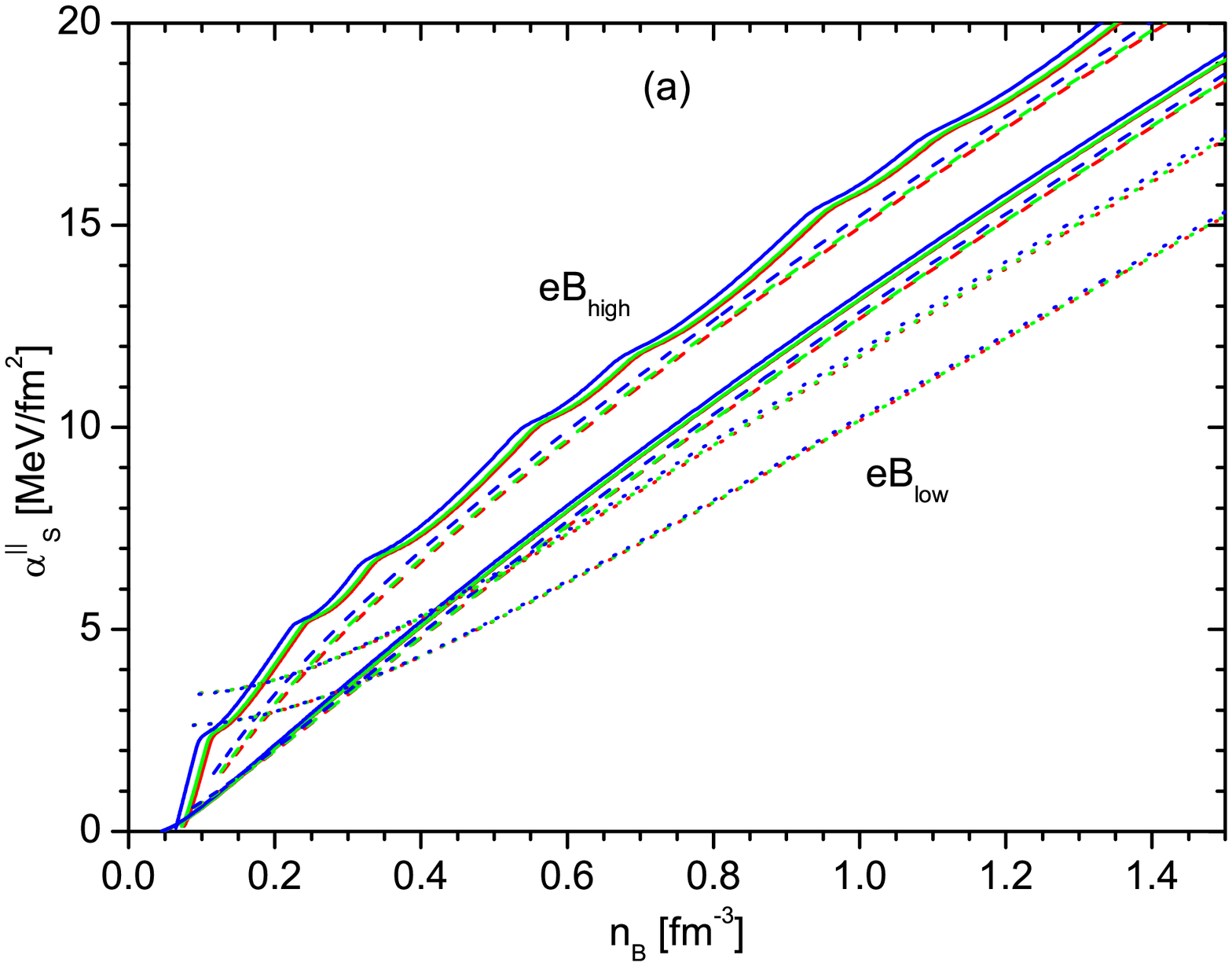}
\includegraphics[angle=0,scale=0.37]{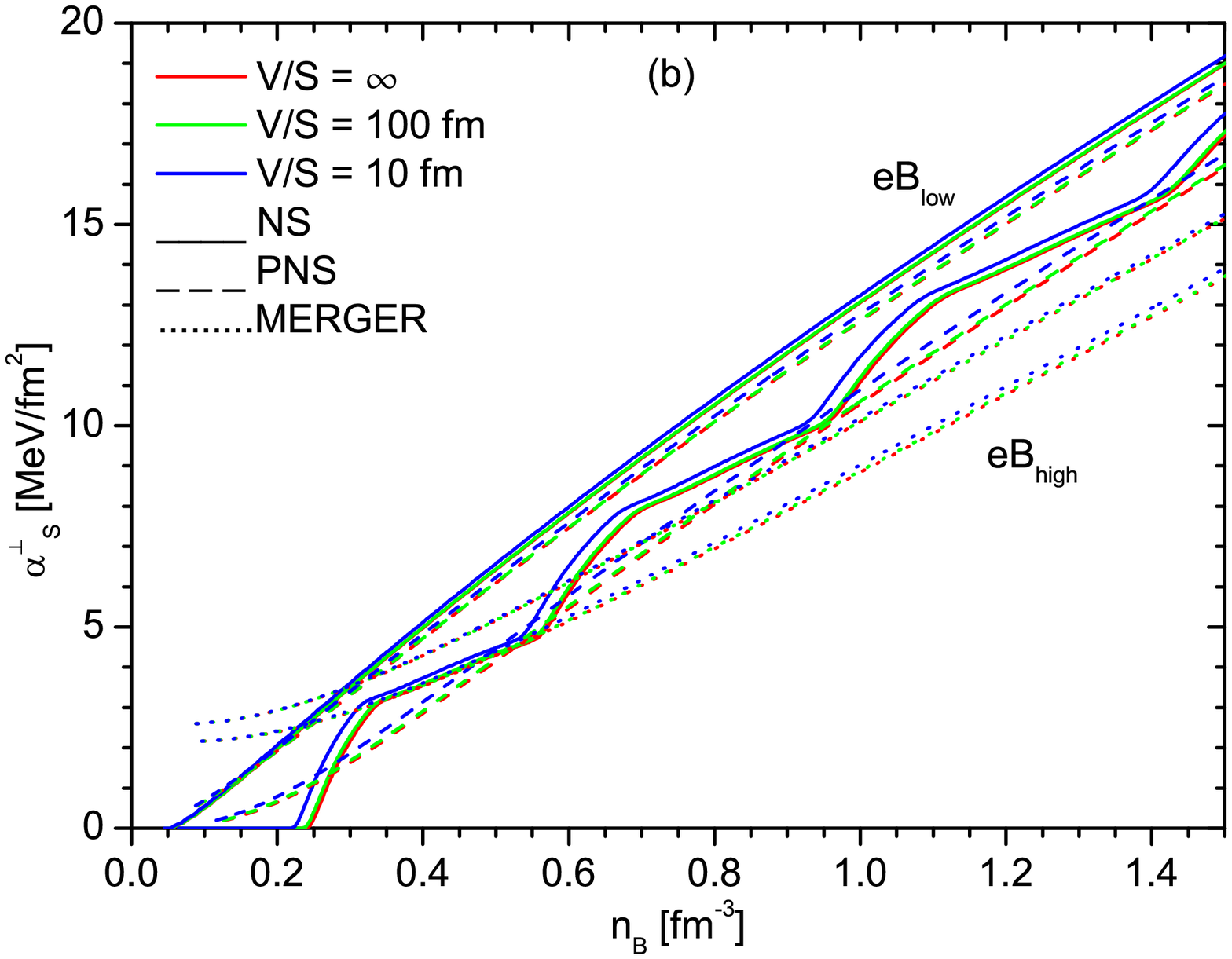}
\caption{Same as in Fig. \ref{fig:1} but for quarks $s$.} 
\label{fig:3}
\end{figure}

\begin{figure}[tb]
\includegraphics[angle=0,scale=0.37]{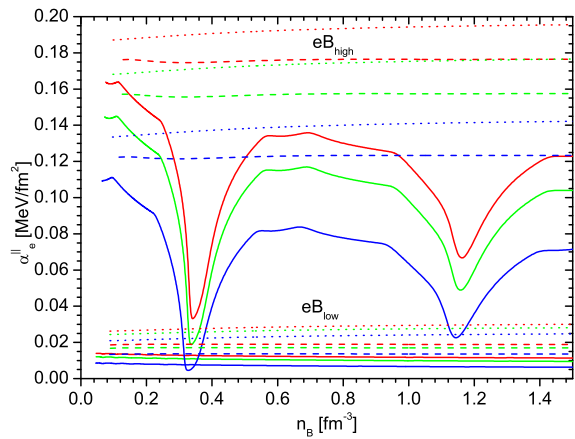}
\caption{Parallel surface tension for electrons. The transverse surface tension is not shown because it is negligible. Lines follow the same pattern as in previous figures.} 
\label{fig:4}
\end{figure}

\begin{figure*}[tb]
\includegraphics[angle=0,scale=0.37]{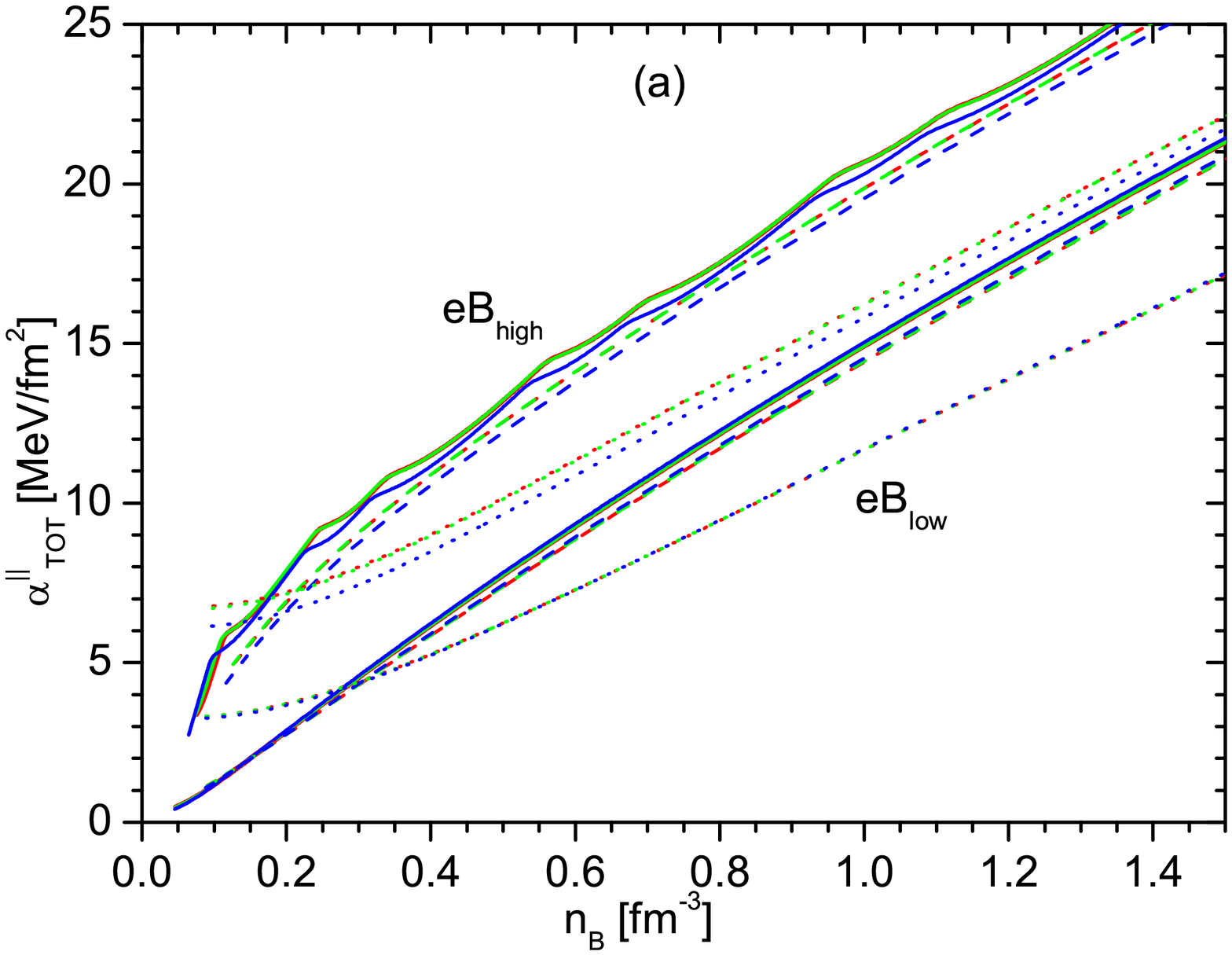}
\includegraphics[angle=0,scale=0.37]{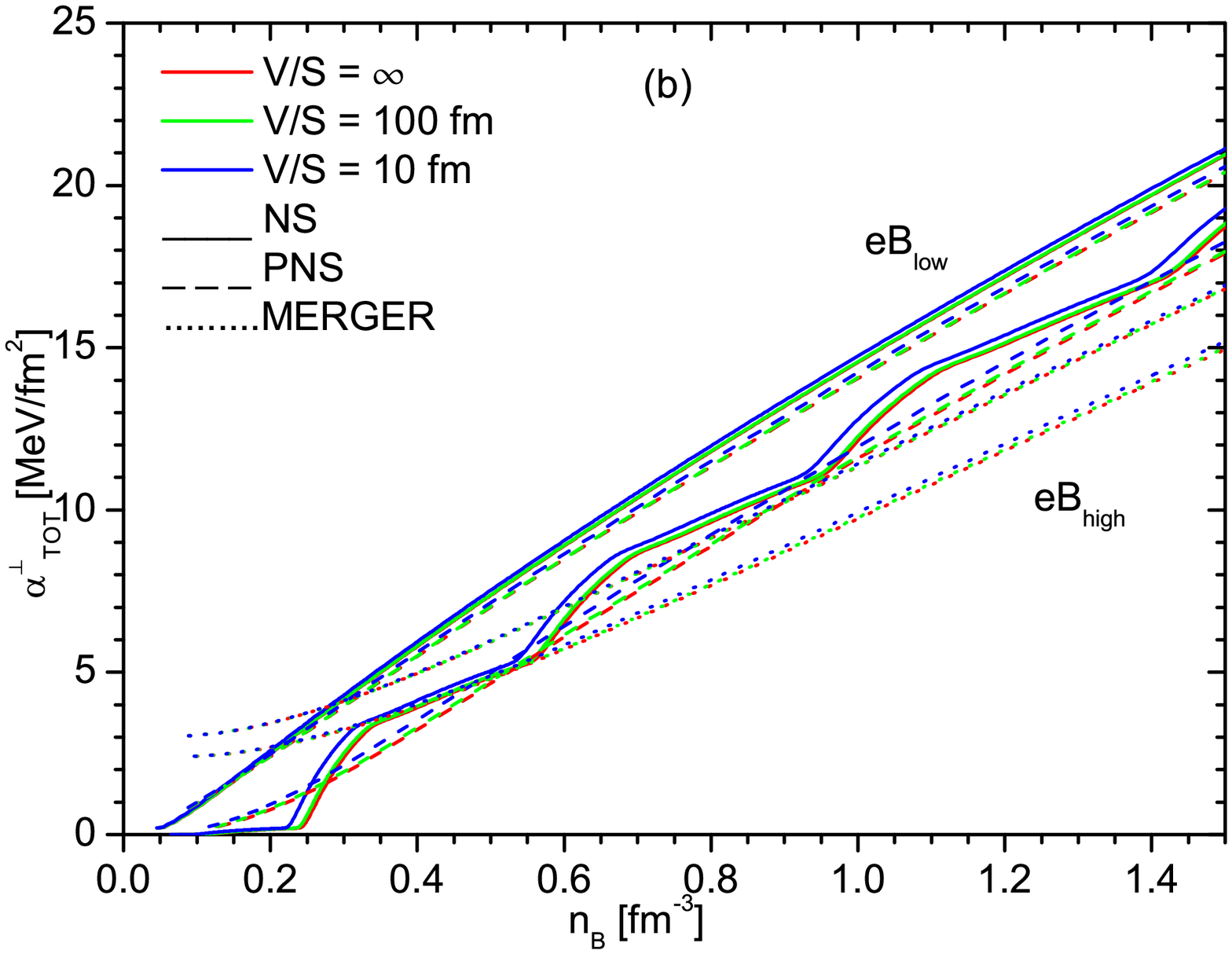}
\caption{Total surface tension obtained by summing the contributions of  quarks $u$, $d$, $s$ and electrons  in the parallel direction (panel (a)) and in the transverse direction  (panel (b)). } 
\label{fig:5}
\end{figure*}

\begin{figure*}[tb]
\includegraphics[angle=0,scale=0.37]{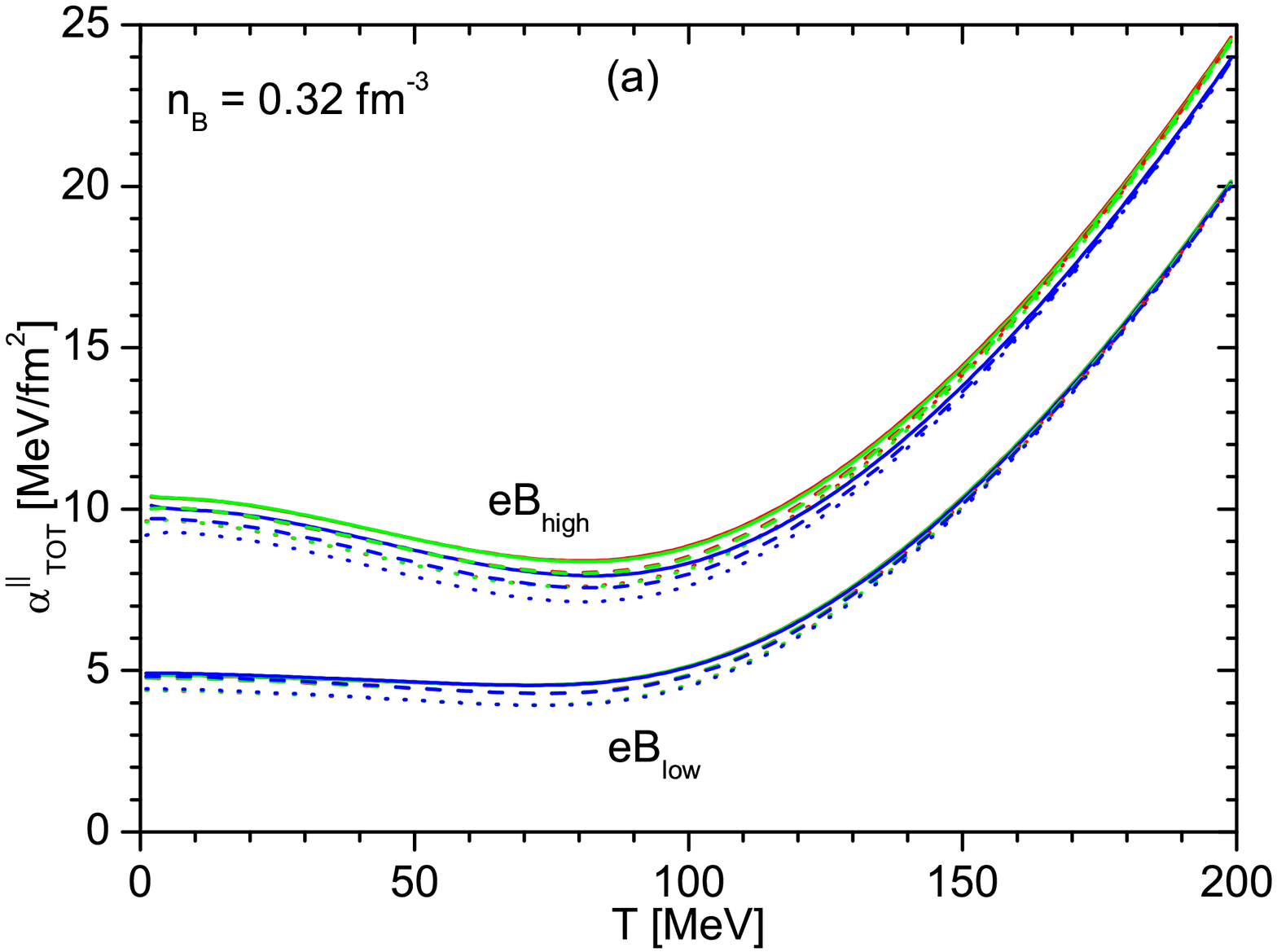} 
\includegraphics[angle=0,scale=0.37]{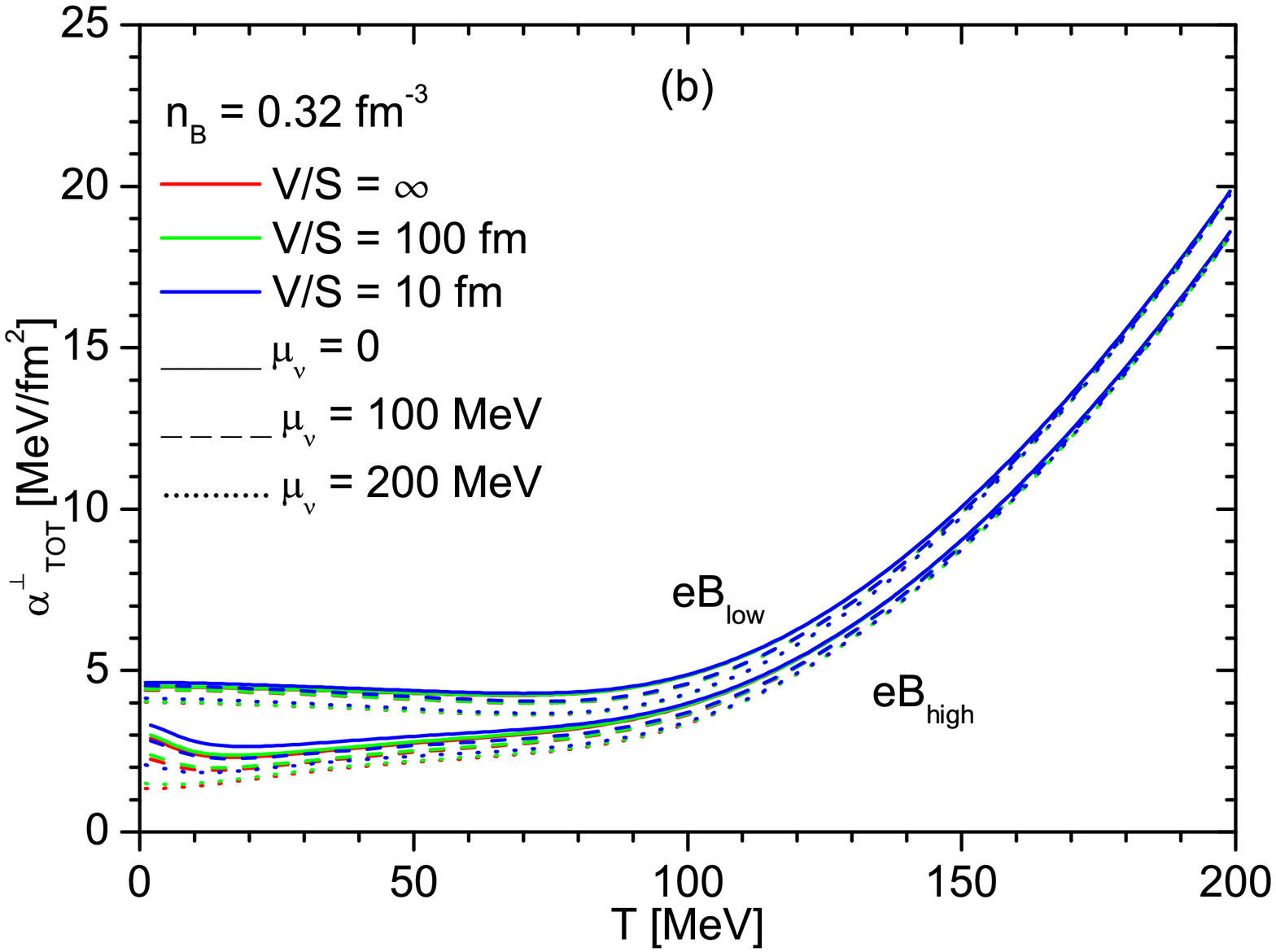}
\includegraphics[angle=0,scale=0.37]{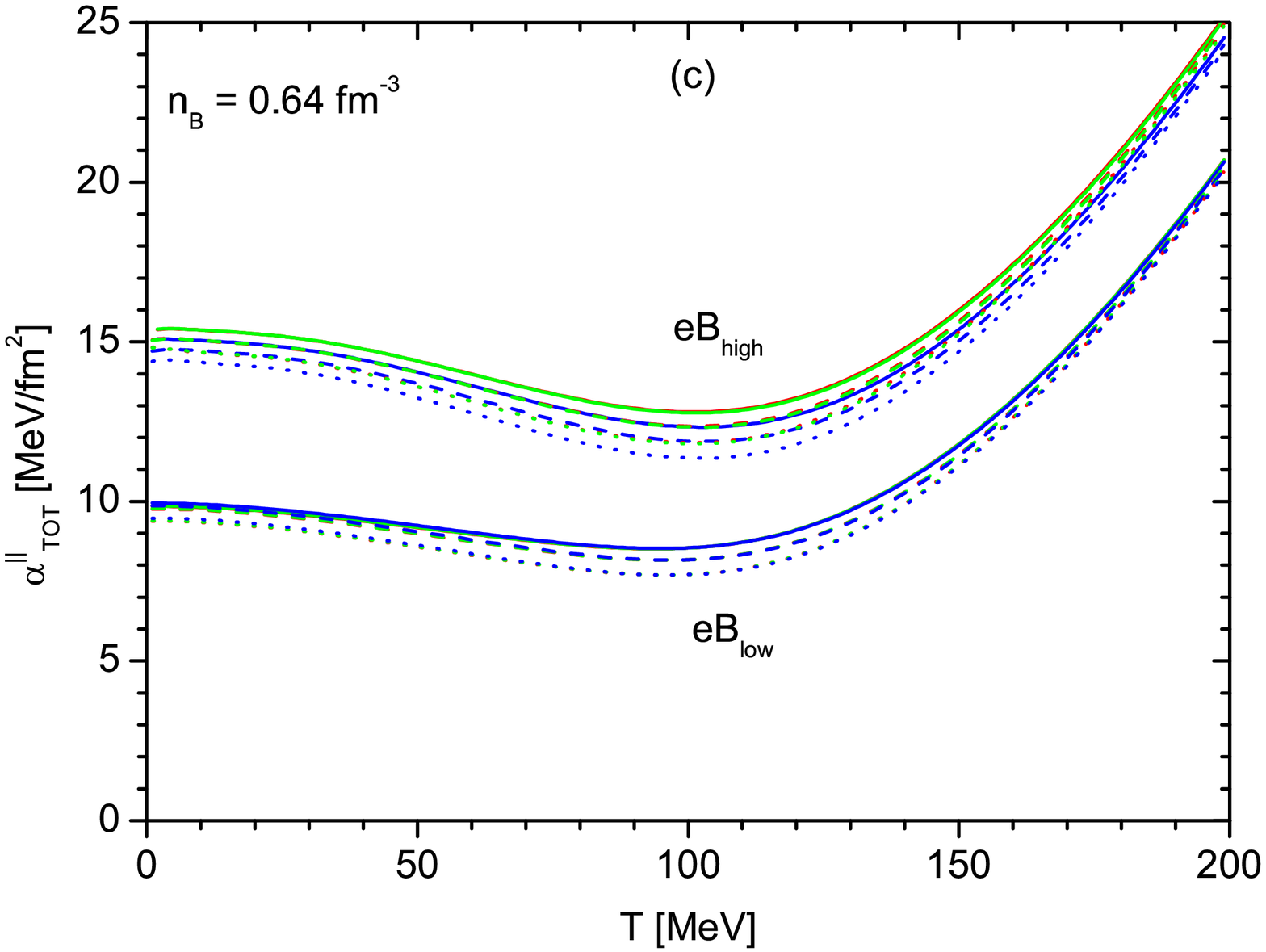} 
\includegraphics[angle=0,scale=0.37]{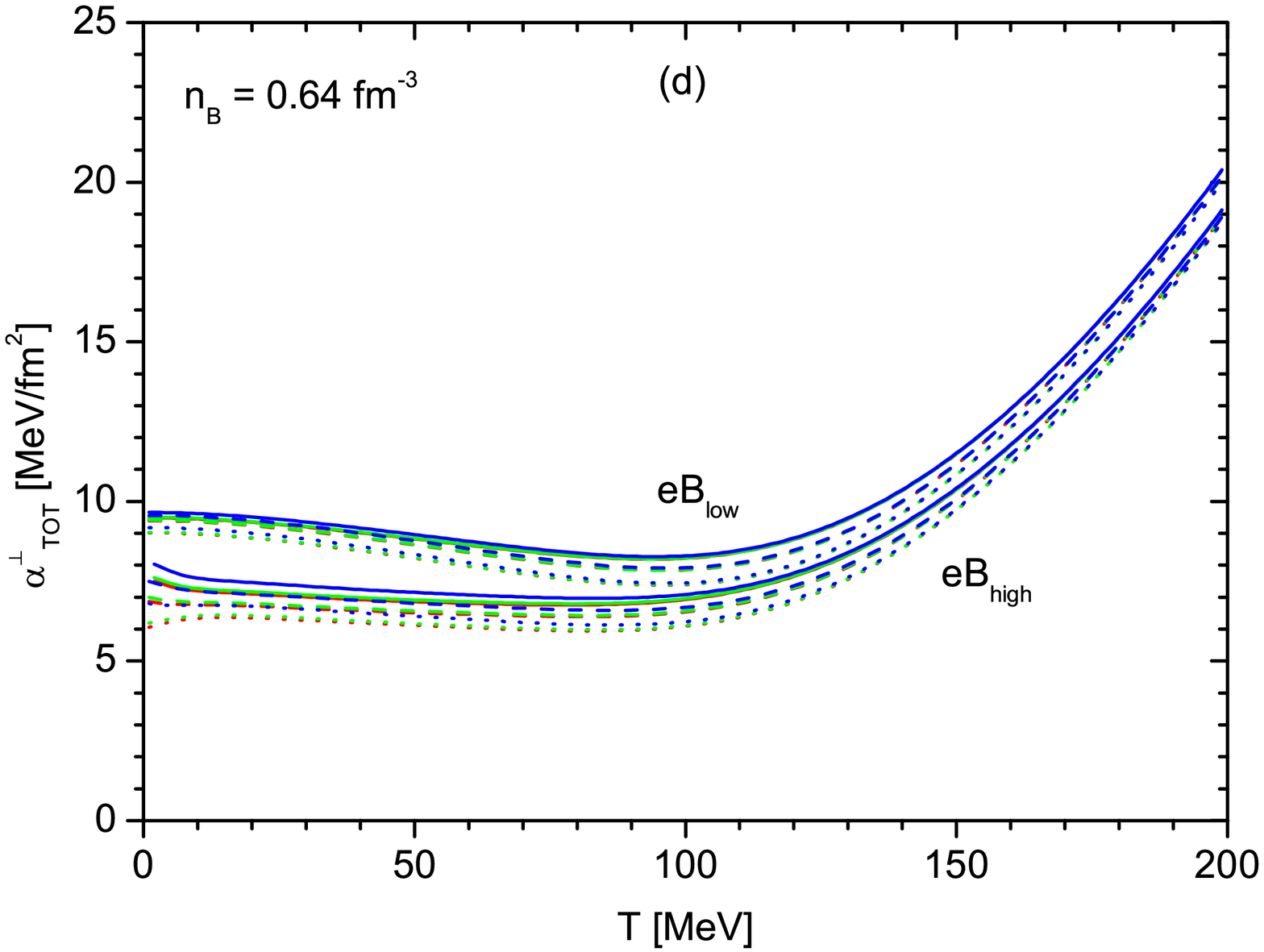}
\caption{Total surface tension obtained by summing the contributions of  quarks $u$, $d$, $s$ and electrons in the parallel direction (panels (a) and (c)) and in the transverse direction  (panels (b) and (d)), as a function of the temperature, for fixed values of the baryon number density ($n_B = 0.32$ and $0.64$ fm$^{-3}$) and the same values of $eB$, $V/S$ and $\mu_{\nu_e}$  used in the previous figures. } 
\label{fig:6}
\end{figure*}

\section{Results}
%
In this section we show our results for the longitudinal ($ \alpha_f^{\parallel}$) and transverse surface tension ($\alpha_f^{\perp}$) focusing on their dependence on the
the baryon number density $n_B = \frac{1}{3} (n_u + n_d + n_s)$, the temperature, the magnetic field strength $eB$, the drop's size $V/S$ and the neutrino chemical potential. 
As emphasized in our previous work \cite{LugonesGrunfeld2017}, $\alpha_f^{\parallel}$ and $\alpha_f^{\perp}$ do not depend on the exact geometry of the drop but only on
the ratio $V/S$, which is taken here as a free parameter (we adopt $V/S = 10$ fm, 100 fm and the bulk limit with V/S = $\infty$).
For the magnetic field strength we consider a low value and a high one: $eB_{low} = 5  \times  10^{-3} \, \mathrm{GeV}^2$ and $eB_{high} = 5  \times 10^{-2}\, \mathrm{GeV}^2$. For magnetic fields smaller than $eB_{low}$ the results are the same as for $eB_{low}$ as already emphasized in our previous work \cite{LugonesGrunfeld2017}.

In Figs. \ref{fig:1}--\ref{fig:4} we show $\alpha_f^{\parallel}$ and $\alpha_f^{\perp}$, for $u,d,s$ quarks and electrons as a function of the baryon number density. For electrons, we show only the longitudinal contribution because the perpendicular contribution is negligible.  
As apparent from Figs. \ref{fig:1}--\ref{fig:3},  for $u$, $d$ and $s$ quarks, the surface tension is an increasing function of the baryon number density, for fixed values of $eB$, $V/S$, $T$ and $\mu_{\nu_e}$. 
In all cases, $\alpha_f^{\parallel}$ is larger for $eB_{high}$  than for $eB_{low}$. For the transverse component the behavior is the opposite, i.e. $\alpha_f^{\perp}$ decreases as the magnetic field increases. 
For very high values of the magnetic field $eB_{high}$, the curves for the NS case clearly show de Haas--van Alphen oscillations related to the filling of new Landau levels. For lower fields, such oscillations are also apparent for the lightest particles. However, for the PNS and MERGER cases, the high temperature allows several excited Landau levels to be fulfilled at any value of the magnetic field. As a consequence, de Haas--van Alphen oscillations are smoothed out.

The effect of varying $V/S$ is minor for heavier particles, like strange quarks (see Fig. \ref{fig:3}). For $u$ quarks  the parallel surface tension for $V/S=10$ fm is decreased with respect to the bulk case by $\sim 10 \%$ for $eB_{low}$ and by $\sim 20 \%$ for $eB_{high}$. For $d$ quarks, the effect is qualitatively the same but quantitatively smaller. The transverse surface tension is almost unaffected by $V/S$  for any value of $eB$. In the case of electrons, the effect of $V/S$ is even larger, because they are lighter.

Now let us focus on the behavior of the surface tension in different astrophysical scenarios. 
In the PNS and MERGER cases the results are affected by neutrino trapping in spite of the vanishing surface tension of the neutrino gas. The influence of neutrinos enters only through the chemical equilibrium condition. 
In the PNS and MERGER cases, the curves are smoother than in the NS case. This is because at finite temperatures the Fermi surface has a thickness of order $T$ which allows Landau levels to be fulfilled more gradually than in the NS case. 
The results for the NS case are in agreement with the results found in our previous work \cite{LugonesGrunfeld2017}. 
For all flavors, $\alpha_{\mathrm{tot}}^{\perp}$ and $\alpha_{\mathrm{tot}}^{\parallel}$ at PNS conditions are smaller than for the NS case, and for the MERGER case they are even smaller.

In Fig. \ref{fig:4} we show the longitudinal surface tension for electrons. The transverse one is negligible because the contribution of the lowest Landau level to $\alpha_{\mathrm{tot}}^{\perp}$ is zero (see Eq. \eqref{eq:transverse_surface_tension}) and at the same time chemical equilibrium leads in general to low values of the electron chemical potential.  
The role of finite volume is significant as well as neutrino trapping and temperature effects. The surface tension decreases with the volume, but increases with temperature, neutrinos and magnetic field.

In Fig. \ref{fig:5} we show the total surface tension in the $\perp$ and $\parallel$ directions.
The surface tension of quark matter in chemical equilibrium under weak interactions is largely dominated by strange quarks (notice that the results in Fig. \ref{fig:5} are very similar to the ones in Fig. \ref{fig:3}). 
The surface tension in the longitudinal and transverse directions is of the order of few MeV/fm$^2$ at baryon number densities around the nuclear saturation density $n_0$ and reaches values of $\sim 15-25$  MeV/fm$^2$ around ten times $n_0$.
The contribution of quarks $u$ and $d$ is always small, with values no higher than 3 MeV/fm$^2$ for the range of baryon number densities of interest
(see Figs. \ref{fig:1}--\ref{fig:2}).  The contribution of electrons is  even smaller and negligible in quark matter, with values below $\sim 0.2$ MeV/fm$^2$ (see Fig. \ref{fig:4}).
In summary, we find that for all the astrophysical scenarios considered here the role of different flavors is qualitatively  the same as already found in Ref. \cite{LugonesGrunfeld2017} for cold deleptonized matter. 

In Fig. \ref{fig:6} we go beyond the three specific astrophysical cases analyzed before and explore  the dependence of $\alpha_{\mathrm{tot}}^{\parallel}$ and $\alpha_{\mathrm{tot}}^{\perp}$ with respect to the temperature for two representative values of the baryon number density, $n_B = 0.32$ and $0.64$ fm$^{-3}$, keeping the same values of $eB$, $V/S$ and $\mu_{\nu_e}$ as before. 
In the range of temperatures below $\sim$ 100 MeV we find that  $\alpha_{\mathrm{tot}}^{\parallel}$  decreases with $T$ while the curves for $\alpha_{\mathrm{tot}}^{\perp}$ are flatter. Above $T \sim$ 100 MeV, the behavior changes  and both $\alpha_{\mathrm{tot}}^{\parallel}$ and $\alpha_{\mathrm{tot}}^{\perp}$ grow significantly with $T$.

\section{Summary and conclusions}

In the present work we have studied the surface tension of hot magnetized quark matter droplets within the formalism of multiple reflection expansion (MRE). Quark matter is described as a mixture of free Fermi gases composed by quarks $u$, $d$, $s$ and electrons in chemical equilibrium under weak interactions. We have considered neutrinos in our system in chemical equilibrium with other particles. Even if they do not contribute explicitly to the surface tension (we considered them to be massless) they do affect the surface tension values through the chemical equilibrium condition. 
Since our system is subjected to the influence of strong magnetic fields the transverse motion of charged particles is quantized into Landau levels, and as a consequence, the surface tension has a different value in the parallel and transverse directions with respect to $B$. We have considered two values of the magnetic field  ($eB_{low} = 5 \times 10^{-3} $ GeV$^2$ and $eB_{high} = 5 \times 10^{-2} $ GeV$^2$) and quark matter drops with sizes given by $V/S$ =  10 fm, 100 fm and  $\infty$. 
We have considered three different astrophysical scenarios depending on the temperature and neutrino chemical potential: cold deleptonized NSs with $T=1$ MeV and $\mu_{\nu_e}=0$,   hot lepton rich PNSs with $T= 30$ MeV and $\mu_{\nu_e} = 100$ MeV, and binary NS mergers with $T= 100$ MeV and $\mu_{\nu_e} = 200$ MeV.

For all these astrophysical scenarios, we find that the dependence of the surface tension with the baryon number density, the magnetic field and the drop's size is essentially the same as found in our previous work at zero temperature and vanishing chemical potential \cite{LugonesGrunfeld2017}. In fact, we find that for $n_B$ between 1 to 10 times the nuclear saturation density, the surface tension falls in the range of $\sim 10^{-2}$ to 25 MeV/fm$^{-2}$ with the larger contribution coming from strange quarks (the most massive particles in the system).
We also find that the surface tensions in the transverse and parallel directions are almost unaffected by the magnetic field if $eB < eB_{low}$, but for higher values of $eB$ there is a significant increase in $\alpha_f^{\parallel}$ and a significant decrease in $\alpha_f^{\perp}$ with respect to the unmagnetized case.
Finally, the \textit{total} surface tension is quite insensitive to the size of the drop, although $V/S$ has a significant effect on the surface tension of light particles (e.g. electrons). 

The main differences with respect to previous results are related to the role of trapped neutrinos and finite temperature.
We found that at fixed baryon number density, magnetic field, $V/S$ and temperature,  both $\alpha_f^{\parallel}$ and  $\alpha_f^{\perp}$ are decreasing functions of the neutrino chemical potential $\mu_{\nu_e}$.  The reduction is less than a few percent, but may be relevant for astrophysics because a smaller surface tension makes easier the nucleation of quark matter in a hadronic star. Thus, the hadron to quark conversion is slightly favored by the large amount of trapped neutrinos in the PNS phase and just after a NS merging. 
We also found that $\alpha_f^{\parallel}$ and  $\alpha_f^{\perp}$ may be significantly affected by temperature.  At fixed baryon number density and fixed neutrino chemical potential we observe that $\alpha_f^{\parallel}$ decreases with $T$ in the temperature range of astrophysical relevance ($T$ below $\sim 100$ MeV) while $\alpha_f^{\perp}$ is less sensitive to temperature.

The combined effect of high temperatures and neutrino trapping can be easily assessed from Fig. \ref{fig:5} where we find that, at fixed baryon number density, the surface tension in the merger case and in the PNS case are always smaller than for cold deleptonized NSs. 
In the PNS case, these results suggest that quark matter nucleation would be more prone to occur in the earlier evolutionary stages of a hadronic star than in later stages. However, as the star cools and deleptonizes, its radius shrinks and the baryon number density in some layers may increase  significantly (typically by a factor of $\sim 2$ \cite{Pons1999}). Contrary to the effect of $T$ and $\mu_{\nu_e}$, this may facilitate quark matter nucleation at late stages. 
Moreover, according to the standard scenario of magnetic field generation in compact stars, the magnetic field inside a PNS could increase during its evolution due to a dynamo mechanism and affect quark matter nucleation if ultrahigh magnetic fields are built inside the object. 
In the merger case, similar effects could be expected, but the theoretical description of their evolutionary stages is still an open question.

\end{document}